# Exploring Metal Additive Manufacturing in Martian Atmospheric Environments

Zane Mebruer, Wan Shou*
Department of Mechanical Engineering, University of Arkansas, AR 72701, USA
*Corresponding author: wshou@uark.edu

**Abstract**

In-space manufacturing is essential for achieving long-term planetary colonization, particularly on Mars, where material transport from Earth is both costly and logistically restrictive. Traditional subtractive manufacturing methods are highly equipment-, energy-, and material-intensive, making additive manufacturing (AM) a more practical and sustainable alternative for extraterrestrial production. Among various AM technologies, selective laser melting (SLM) stands out due to its exceptional versatility, precision, and capability to produce dense metallic parts with complex geometries. However, conventional SLM processes rely heavily on inert argon environments to prevent oxidation and ensure high-quality part formation—conditions that are difficult to reproduce on Mars. This study investigates the feasibility of using carbon dioxide ($CO_2$)—which makes up over 95% of the Martian atmosphere—as a potential substitute for argon in SLM. Single-track and two-dimensional 316L stainless steel specimens were fabricated under argon, $CO_2$, and ambient air environments with a wide range of laser parameters to evaluate the influence of atmospheric composition on surface morphology, microstructural cohesion, and oxidation behavior. The results reveal that no single parameter controls the overall part quality; rather, a balance of parameters is essential to maintain thermal equilibrium during fabrication. Although parts produced in $CO_2$ exhibited slightly inferior surface finish, cohesion, and oxidation resistance compared to argon, they performed significantly better than those fabricated in ambient air. These findings suggest that $CO_2$-assisted SLM could enable sustainable in-situ manufacturing on Mars and may also serve as a cost-effective alternative shielding gas for terrestrial applications.

## 1. Introduction

Establishing a sustainable human presence on Mars presents immense scientific, engineering, and logistical challenges. Among these, one of the most critical hurdles is to overcome cargo constraints associated with interplanetary transport. Launching materials from Earth is extremely expensive and limited by strict mass and volume restrictions, making it impractical to continuously supply all the tools, components, and replacement parts needed to sustain a long-term mission. To ensure the success and longevity of Martian colonization, it will therefore be essential to develop in-situ resource utilization (ISRU) and in-situ manufacturing capabilities that allow astronauts to fabricate and repair equipment directly on the Martian surface [1]. Traditional subtractive manufacturing methods—such as machining and milling—have proven indispensable on Earth for producing precise and reliable components. However, these processes are energy-intensive, material-wasteful, and dependent on heavy machinery that is impractical to transport to extraterrestrial environments. In contrast, additive manufacturing (AM), commonly referred to as 3D printing, offers an attractive alternative for space applications [2]. AM allows the creation of complex geometries directly from digital designs with minimal material waste, reduced energy consumption, and lower system mass—key advantages under the severe logistical constraints of deep-space missions.

Among the various AM technologies, light-based selective laser melting (SLM) stands out as one of the most versatile and widely adopted techniques for metal-based fabrication [3]. In the SLM process, fine metal powder layers are sequentially melted by a focused laser beam to produce fully dense, near-net-shape components. However, the process is typically conducted within an inert gas environment—most commonly argon—to prevent oxidation and other detrimental chemical reactions that can impair surface finish and degrade mechanical performance. Yet, establishing and maintaining an argon atmosphere on Mars would pose significant challenges, as argon is scarce and would have to be transported from Earth or extracted in small quantities from the Martian atmosphere. Interestingly, carbon dioxide ($CO_2$) constitutes more than 95% of the Martian surface atmosphere [4-7] and is chemically stable under moderate temperature conditions. Although $CO_2$ is not a noble gas, its potential to act as a processing medium under a controlled environment for SLM has never been studied. If metal AM could be effectively conducted within a $CO_2$ atmosphere, this would represent a major advancement toward self-sustaining manufacturing infrastructure on Mars, dramatically reducing the need for imported gases and a protected chamber. In other words, if $CO_2$ does not affect the processing, AM can be conducted in its atmospheric condition without dimensional constraints.

In this work, we use $CO_2$ filled chamber to simulate the Mars atmosphere environment [7] for SLM of a widely used metal, 316L stainless steel [8-10], as an important model material. Through single-track and layer-based fabrication experiments, we systematically examine the influence of atmospheric composition on melting, surface morphology, and oxidation behavior. By comparing results obtained under argon, $CO_2$, and ambient air environments, this study provides fundamental insights into the role of process atmosphere in metal AM and evaluates the potential of $CO_2$-assisted SLM for future extraterrestrial manufacturing. Beyond its relevance for space applications, the findings also have implications for terrestrial manufacturing, where $CO_2$ may offer a more sustainable and cost-effective alternative to traditional inert gases.

2. **Materials and Methods**

316L steel with an average particle size between 22 – 28 μm was used as the testing material due to its compliance with SLM requirements, handling simplicity, and applicability for Martian colony needs [11]. The manufacturing system (as shown in Figure 1a) comprises a fiber laser with a wavelength of 1064 nm and a maximum power of 80 W, and a customized power bed (Figure 1b). This customized chamber has a quartz window, a gas inlet and outlet, which allows us to simulate the desired atmosphere environment. The specific components to construct the chamber are shown in Figure 1d. The fabrication chamber was simple to meet the basic requirements for SLM with a controllable gas environment. The main body of the chamber was composed of two aluminum plates that were machined via CNC mill to create an open-top chamber when the two pieces were put together. A quartz glass window was installed for the viewing port through which the laser shone. To create an air-tight seal, a gasket cut to shape from a rubber stamp sheet was implemented between the two aluminum plates. The quartz window was attached to the chamber with an air-tight seal. The atmosphere of the chamber was controlled via two valves installed in the chamber. One valve was used to pump in the desired atmosphere, while the other was used as an exit valve. The two chamber halves were held together via four bolts. The metal powder was contained within the chamber in a small 3D printed powder bed (as shown in Figure 1c).

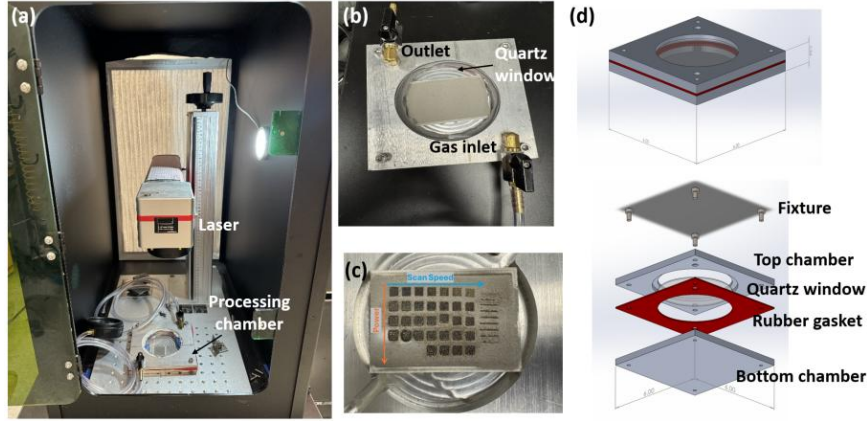

**Figure 1** Experimental setup for selective laser melting with an artificial environment.

Due to the window's proximity to the metal powder and the high energy being focused onto the powder, small amounts of powder would vaporize and deposit on the window's inner surface. Therefore, a glass slide was attached to the bottom of the quartz window to protect the window. A new glass slide was used after each manufacturing batch.

A repeatable method for preparing samples was established to achieve optimal project efficiency and ensure repeatability. First, the 316L metal powder was distributed in the powder bed. Then, the powder bed was leveled with a glass slide. The powder bed was then placed into the manufacturing chamber. If an atmosphere other than ambient was tested, one of the chamber vents was then secured with a vacuum gauge to verify proper chamber seal. The other was attached to a one-stage vacuum pump that was allowed to run for 30 seconds to ensure proper ambient atmosphere removal. The vacuum pump was then disconnected, and the chamber was connected to either the $CO_2$ or Ar tank. Before connection, the line between the chamber and the gas tank was ensured to be free of ambient atmosphere. Once connected, the gas was allowed to slowly fill the chamber until the vacuum gauge was at atmospheric pressure. The chamber was then disconnected and placed below the laser.

Both single-track and two-dimensional (2D) samples were fabricated. For single-track samples, laser powers ranged from 16 – 80 W, at scanning speeds from 10 – 200 mm/s, and frequencies from 10 to 4000 kHz. 2D samples were manufactured at powers ranging from 16 – 48 W, scanning speeds from 50 – 200 mm/s, laser frequencies of 500 kHz, and hatch spacings of 20 and 35 μm. Laser parameters, batch layouts, and sample patterns were controlled via Lightburn software.

All optical images were taken via a Tomilov digital microscope. ImageJ software was utilized to analyze the dimensions of the manufactured samples. 2D samples were compared using the area sintering percentage to understand the feasibility and accuracy for square sample fabrication. The area of a given sample was measured via ImageJ area measurement tools and was compared to the total target sintering area of the sample. The target sample area was determined by etching a square on wood with the chosen Lightburn size parameter of $5 \times 5$ mm$^2$. The square was then measured with digital calipers to determine the target area, which was defined as 26.884 mm$^2$.

Part cohesion of 2D was evaluated through an area retention calculation, which was referred to as solidification percentage:

$$Solidification\ percentage = \frac{Calculated\ Sample\ Area}{Target\ Area} \times 100\%$$

Selected samples were further characterized using a FEI Nova Nanolab 200 Scanning Electron Microscope (SEM) to obtain high-resolution microstructural images and assess surface morphology and quality. Besides, Energy-Dispersive X-ray (EDX) Spectroscopy was employed to evaluate compositional differences and identify oxidation-related byproducts across the tested atmospheres. While the EDX measurements were not intended to provide quantitative elemental compositions, they offered valuable comparative insights into the relative oxygen presence and surface chemistry among samples fabricated under $CO_2$, argon, and ambient conditions.

## 3. Results and Discussions
### 3.1 Single-track sintering

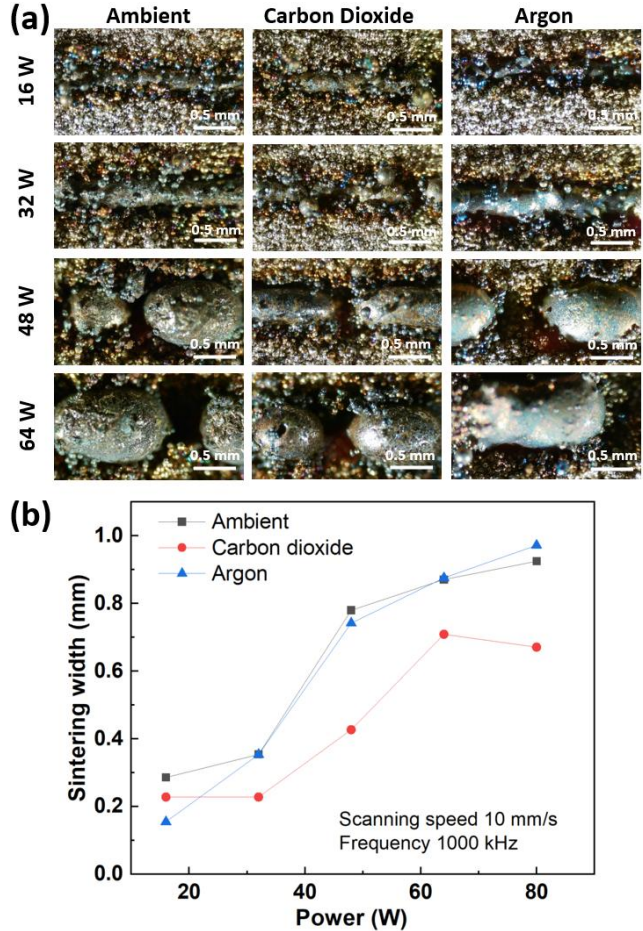

**Figure 2** (a) Laser power effect on single-track sintering (1000 kHz, 10 mm/s); (b) line width change with power.

To understand the influence of laser parameters (including power, frequency and scanning speed) on the sintering results, we first conduct single-track scanning of metal powder with three different atmosphere environments (ambient, Martian atmosphere (namely, carbon dioxide), and argon). Morphology changes are obvious with different laser powers (as shown in Figure 2a). Generally, when the power is increased from 16 to 80 W, the line width significantly increases regardless of the processing environment (Figure 2b). With relatively low power, continuous lines can be fabricated; however, with higher powers (≥48 W), the balling effect becomes prominent [12], forming discontinuous, elongated sections. The correlation between laser power and single-track line width is summarized in Figure 2b. The general trend is that line width increases as laser power, or more accurately, linear energy density (LED, defined as $\frac{P}{v}$, where $P$ denotes the laser power, and $v$ denotes the scanning speed [13]). Compared with samples fabricated in ambient and argon environments, the $CO_2$ environment tends to give the smallest line width, which might be attributed to the reduced interfacial energy between $CO_2$ and molten stainless steel, and accompanied reaction [14,15].

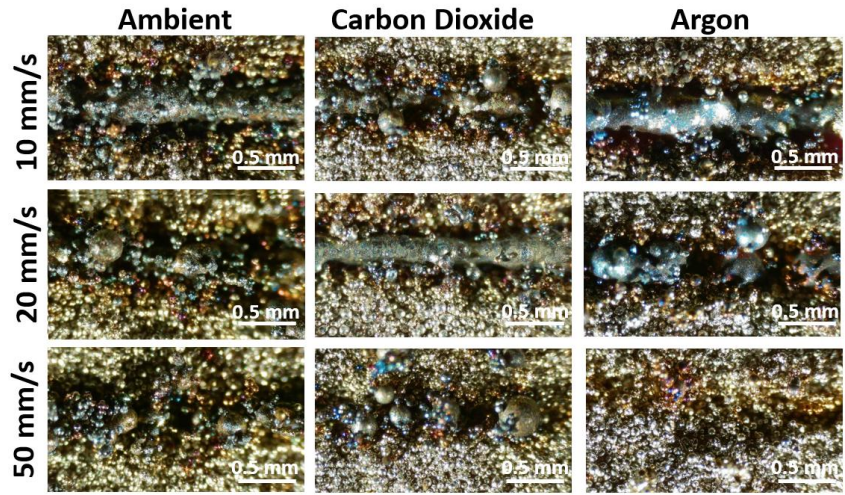

**Figure 3** Scanning speed effect on single-track samples (1000 kHz, 32 W).

We further studied the influence of scanning speed on the sintered morphology (with a fixed power of 32 W). As shown in Figure 2, with the increase of scanning speed from 10 to 50 mm/s (or decrease of linear energy density), the sintered line becomes discontinuous. In comparison of different processing environments, the argon environment gives the widest continuous line, followed by the ambient environment, and then the $CO_2$ environment. Yet, it is noticed that with the $CO_2$ environment, both 10 mm/s and 20 mm/s can form continuous lines, while the ambient and argon environment only forms continuous lines at 10 mm/s. This suggests that with different processing environments, the molten stainless steel may have different surface tension on the solid powder bed.

**3.2 Fabrication of 2D samples**

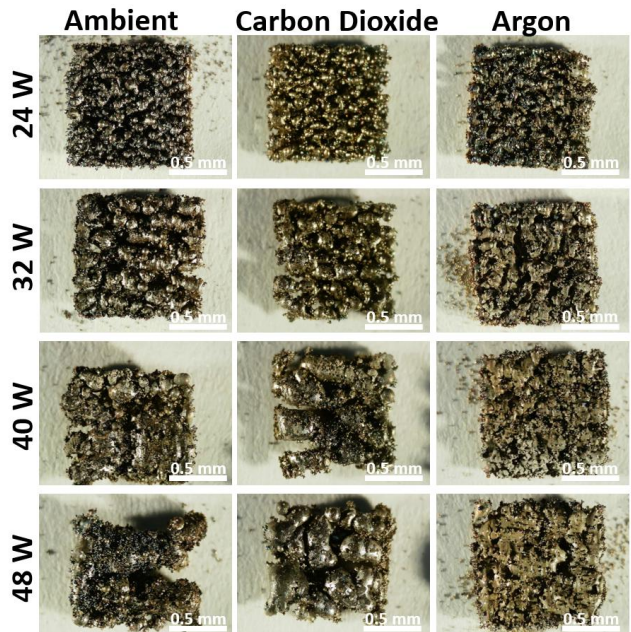

**Figure 4** Laser power effect on fabricated 2D samples (125 mm/s, 20 μm hatch space).

Following single-track sintering, we explored the fabrication of 2D samples in different environments with a variety of parameter combinations. Generally, with the increase of laser power, the balling effect becomes more obvious (as shown in Figure 4), especially for ambient and $CO_2$ environments. Samples fabricated with 24 W contain much smaller beads of melted powder than the ones made with higher powers. By increasing the laser power to 32 W, the melting and coalescence become more evident as the heat accumulates in the powder bed, forming a connected 2D plate, especially the one with argon. Generally, with relatively low power (≤32 W or LED, 0.256 J/mm), samples show better shape retention. Clear cracks were observed in the samples fabricated with power ≥40 W (or LED≥0.32 J/mm) in ambient and $CO_2$ environments, which is caused by overheating and thermal stress. Yet, samples fabricated with an argon environment have better cohesion, which suggests that the oxidation of stainless steel in ambient and $CO_2$ reduces the effective coalescence.

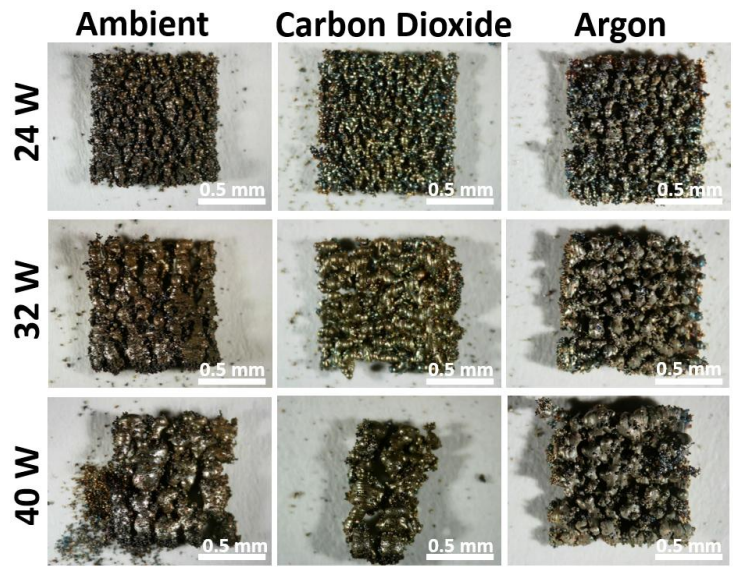

**Figure 5 Laser power effect on fabricated 2D samples (125 mm/s, 35 μm hatch space).**

Based on these observations, we further increased the hatch space from 20 μm to 35 μm, using laser power between 24-40 W. It is observed that with a smaller hatch space and larger overlapping area, the sintered structures tend to be more cohesive and have fewer gaps between different tracks. When the hatch space is 35 μm, 40 W laser power, unfortunately, can not sinter the powders into a whole piece in an ambient and $CO_2$ environment. This failure may be caused by the over-oxidization and insufficient heat to bond the melts together.

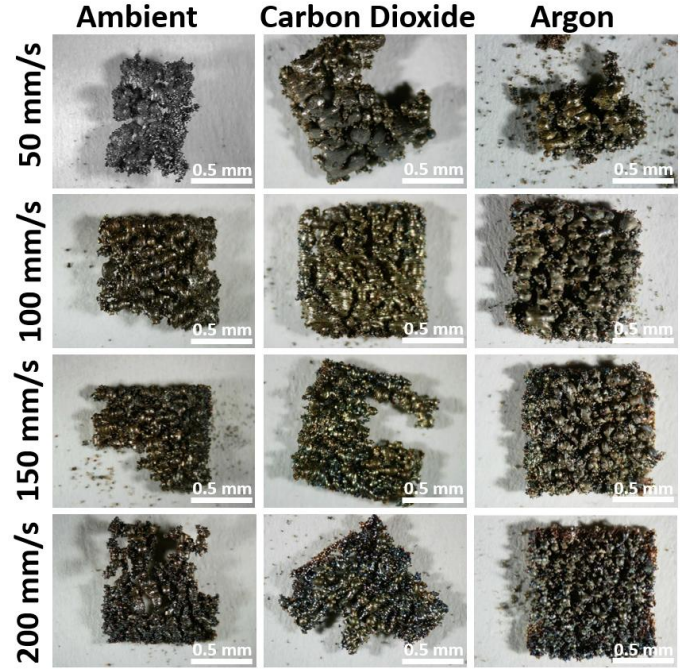

**Figure 6** Scanning speed effect on 2D samples (32 W, 35 μm hatch space).

Compared with samples with 20 μm hatch space and 32 W, we feel 35 μm hatch space tends to retain a better shape. Thus, we chose a 35 μm hatch space and fixed the laser power at 32 W to further check the influence of scanning speed on the morphology of 2D samples. As shown in Figure 6, by reducing scanning speed from 125 mm/s to 100 mm/s (or increasing LED to 0.32 J/mm), the resulting morphology is similar to the cases in Figure 4 with a power of 40 W (corresponding to a LED of 0.32 J/mm). Yet, further decrease in scanning speed to 50 mm/s (or double the LED to 0.64 J/mm), the designed shape is no longer retained (first row in Figure 6). On the other hand, an increase in scanning speed from 125 mm/s to 150 mm/s and 200 mm/s (corresponding to lower LED of 0.213 J/mm and 0.16 mm/s) results in partially sintered pieces in ambient and $CO_2$ environments. However, with argon protection, high speed (or low LED) can still form a well-shaped piece. Overall, the argon environment is suitable for the widest range of scanning speeds, while the $CO_2$ environment works better than the ambient environment for SLS of stainless steel powder.

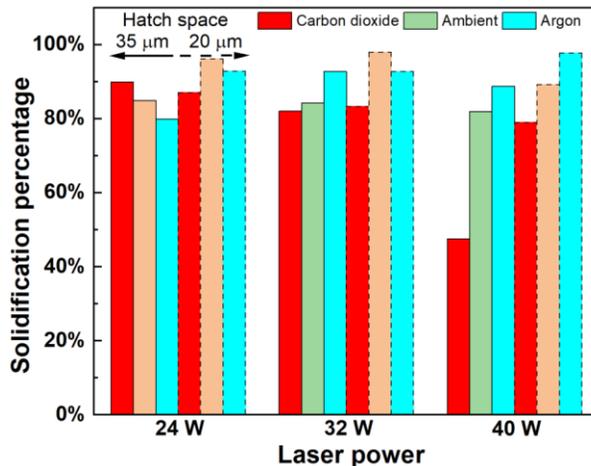

**Figure 7** Solidification percentage with different hatch spaces.

Lastly, we quantified all the samples with solidification percentage, and illustrated them in Figure 7. Generally, a balanced linear energy density can lead to a more consistent solidification percentage with the designed shape. The graph shows that a 20 µm hatch spacing retains its area more effectively than a 35 µm spacing across all tested laser powers at a scanning speed of 125 mm/s, indicating that optimized hatch spacing is critical to maintaining the structural integrity of the fabricated part. The inert gas (i.e., argon) seems to be the most suitable (and most widely used) processing environment for stainless steel, followed by a simulated Martian environment (i.e., $CO_2$ gas), and then the ambient. These observations suggest that the Martian atmospheric environment (i.e., $CO_2$ gas) can potentially be used directly for selective laser sintering/melting.

### 3.3 Atmospheric effect on morphology and oxidation

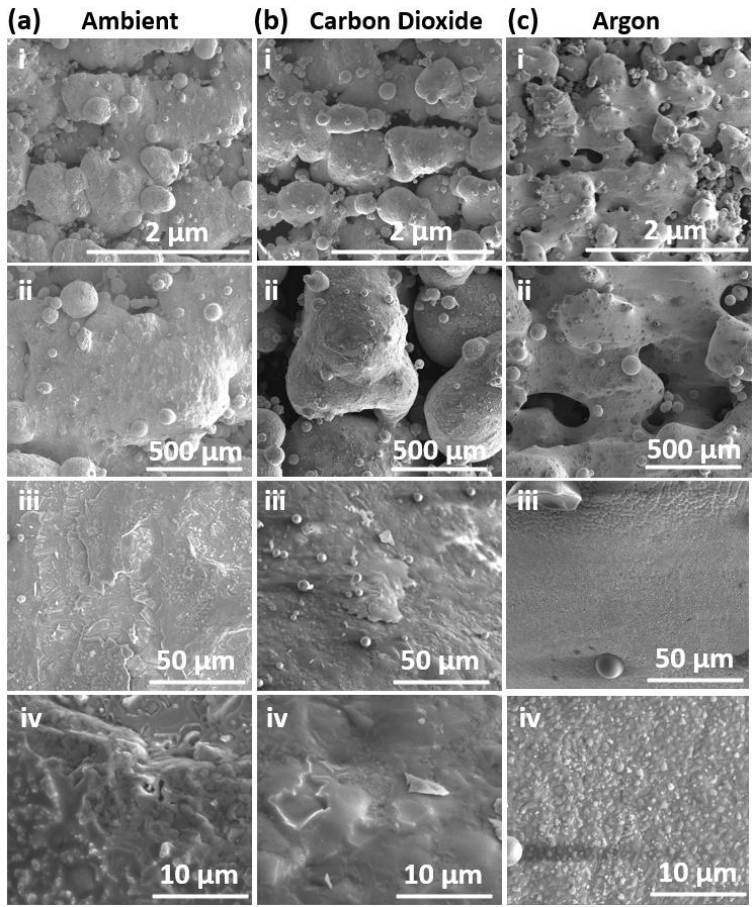

**Figure 8** SEM image of samples fabricated in three different atmospheres with scanning speed of 125 mm/s, power of 32 W, and hatch space of 35 µm.

To further assess the potential oxidation effect on the fabricated samples, 2D samples processed under each atmospheric condition were characterized by SEM and EDX. As indicated in Figure 8, consistent with earlier observations, samples processed in an argon environment exhibit the most uniform and dense surfaces (Figure 8c), whereas processing in ambient air (Figure 8a) and $CO_2$ (Figure 8b) leads to progressively higher levels of oxidation-related defects and surface roughness. Across all magnifications, argon-processed samples display significantly smoother and more continuous morphologies (Figure 8c). At low magnification, uniform melt tracks with minimal balling are observed. At higher magnifications, the surface appears dense and fine-textured, with only isolated droplets. The inert argon atmosphere effectively suppresses oxidation during laser processing (indicated in Table 1, with the lowest atomic percentage of oxygen), enabling stable melt pools, improved wetting, and enhanced inter-track fusion.

In contrast, samples processed in ambient air (Figure 8a) and $CO_2$ (Figure 8b) exhibit pronounced melt instabilities. At low magnification, both environments show a substantial balling effect (Figure 8, row i), characterized by numerous spherical and semi-spherical features, indicating unstable melt flow during laser irradiation. For ambient-air processing, intermediate-magnification images reveal irregular and uneven surfaces with resolidified droplets distributed along the melt tracks. At higher magnification (Figure 8, row iv), rough surface textures and microscale protrusions are evident. The presence of oxygen promotes oxidation during laser melting, enhancing surface-tension gradients and inhibiting smooth melt flow; oxide formation further suppresses wetting, leading to balling and poor inter-track bonding. The high atomic percentage of oxygen in the ambient-processed sample was confirmed by EDX analysis, as shown in Table 1. Compared to ambient air, $CO_2$ processing results in larger and more agglomerated melt features at low magnification (Figure 8b-i). At intermediate magnification, irregular solidified lumps and partially fused regions are observed, while high-magnification images reveal uneven surfaces with resolidification ridges and fine microparticles. Although $CO_2$ reduces oxygen availability relative to ambient air, thermal decomposition or residual oxygen still facilitates surface reactions. A lower percentage of oxygen was detected in the sample processed in the simulated Martian atmospheric environment (as shown in Table 1). This may increase the melt viscosity and flow instability, resulting in coarser surface features compared with the sample processed in argon.

**Table 1** EDX analysis of the sample processed in different environments.

| Element | Atomic percentage (%) | | |
|---|---|---|---|
| | Ambient | $CO_2$ | Argon |
| Carbon | 17.59 | 15.11 | 17.10 |
| Oxygen | 53.76 | 44.60 | 27.68 |
| Chromium | 14.17 | 11.21 | 11.92 |
| Manganese | 11.65 | 3.05 | 1.13 |
| Iron | 2.61 | 22.23 | 36.95 |
| Nickel | 0.21 | 3.43 | 4.46 |
| Molybdenum | 0.01 | 0.37 | 0.75 |

## 4. Conclusions

Single-track and two-dimensional samples were manufactured using SLM with 316L steel in argon, carbon dioxide, and ambient atmospheric environments to investigate the influence of atmosphere on the morphology. It was found that the optimal morphology is less dependent on the atmospheric environment, compared with laser power and scanning speed, as well as hatch space for 2D specimens. If thermal effects are not strong enough, the powder would see insufficient melting and not form a cohesive bond. If thermal effects are too strong, balling effects would occur and create metal spheres as a result of capillary forces. It was found that argon was the superior fabrication atmosphere over either carbon dioxide or ambient. Samples created in argon had consistently better cohesion, area retention, and surface quality. Argon also exhibited lower amounts of oxidation than other atmospheres. However, carbon dioxide did exhibit better performance than the ambient atmosphere in terms of cohesion, which is promising for potential SLM on Mars.

Further work is needed to verify the suitability of a carbon dioxide atmosphere for SLM. Three-dimensional components should be fabricated and tested to quantify how surface quality, adhesion, and oxidation influence mechanical performance. Studies involving other alloys and materials will help clarify how composition affects process outcomes. Future research should also explore integrating $CO_2$-based solar-driven additive manufacturing concepts [16,17] to enable energy-efficient, in-situ fabrication on planetary surfaces—an important step toward autonomous manufacturing for space exploration.

## Acknowledgments

This work was supported by the Honors College Undergraduate Research Grant and the Startup Package from the University of Arkansas.